\documentstyle[psfig]{nature_local}

\def\ltsima{$\; \buildrel < \over \sim \;$}
\def\simlt{\lower.5ex\hbox{\ltsima}}
\def\gtsima{$\; \buildrel > \over \sim \;$}
\def\simgt{\lower.5ex\hbox{\gtsima}}

\title{Rapid Growth of Black Holes in Massive Star-Forming Galaxies}

\author{D.~M.~Alexander\affiliation{Institute of Astronomy, Madingley Road, Cambridge CB3 0HA, UK},
        I.~Smail\affiliation{Institute for Computational Cosmology, University of Durham, South Road, Durham DH1 3LE, UK},
        F.~E.~Bauer $^{*}$,
        S.~C.~Chapman\affiliation{California Institute of Technology, Pasadena, CA 91125, USA},
        A.~W.~Blain $^{\ddag}$,
        W.~N.~Brandt\affiliation{Department of Astronomy and Astrophysics, Pennsylvania State University, 525 Davey Laboratory, University Park, PA 16802, USA},
        R.~J.~Ivison\affiliation{Astronomy Technology Centre, Royal Observatory, Blackford Hill, Edinburgh EH9 3HJ, UK}\affiliation{Institute for Astronomy, University of Edinburgh, Blackford Hill, Edinburgh EH9 3HJ, UK}
}

\headertitle{}
\mainauthor{}

\summary{\noindent The tight relationship between the masses of black
  holes and galaxy spheroids in nearby galaxies$^{1}$ implies a causal
  connection between the growth of these two components. Optically
  luminous quasars host the most prodigious accreting black holes in
  the Universe and can account for $\simgt30$~\% of the total
  cosmological black-hole growth$^{2,3}$. As typical quasars are not,
  however, undergoing intense star formation and already host massive
  black holes ($>10^{8}$~$M_{\odot}$)$^{4,5}$, there must have been an
  earlier pre-quasar phase when these black holes grew (mass range
  $\approx$~$10^{6}$--$10^{8}$~$M_{\odot}$). The likely signature of
  this earlier stage is simultaneous black-hole growth and star
  formation in distant (i.e., $z>1$; $>8$~billion light years away)
  luminous galaxies. Here we report ultra-deep X-ray observations of
  distant star-forming galaxies that are bright at submillimetre
  wavelengths. We find that the black holes in these galaxies are
  growing almost continuously throughout periods of intense star
  formation. This activity appears to be more tightly associated with
  these galaxies than any other coeval galaxy populations. We show
  that the black-hole growth from these galaxies is consistent with
  that expected for the pre-quasar phase.}  \dates{12 October 2004}{11
  February 2005}

\begin{document}
\maketitle

The most intense sites of star formation at high redshift are
associated with submillimetre galaxies (SMGs)$^{6-9}$. SMGs are
amongst the most bolometrically luminous galaxies in the Universe
($L_{\rm BOL}\approx10^{13}$~$L_{\odot}$) and the majority of the
population is coeval with the peak in quasar activity (i.e.,
$z\approx$~1.5--3.0)$^{7,9}$. The apparent association of SMGs with some
quasars and the similarity in the comoving space densities of SMGs and
optically luminous ($M_{\rm B}<-24$) quasars (when corrected for
probable source lifetimes) provides direct evidence for an
evolutionary connection between SMGs and quasars$^{5,10,11}$. However,
in contrast to quasars, the bolometric output of SMGs appears to be
dominated by powerful star-formation activity and any black-hole
accretion [i.e., Active Galactic Nuclei (AGN) activity] is
comparatively weak$^{12,13}$. Given the large available molecular gas
supply (typically $\approx3\times10^{10}$~$M_{\odot}$)$^{14}$, SMGs
can fuel this luminous star-formation activity for $\approx10^{8}$~yrs
(ref 14,15). Various pieces of complementary observational support
have also shown that SMGs are massive galaxies (typically
$\approx10^{11}$~$M_{\odot}$), suggesting that they will become
$\simgt$$M_{*}$ spheroid-dominated galaxies in the local
Universe$^{14,16,17}$.

AGN activity has been identified in many SMGs$^{6,12}$. However, the
difficulty in obtaining high-quality optical spectra and reliable
source redshifts has hindered a complete census of AGN activity in
SMGs. We have initiated a project investigating the properties of AGNs
in SMGs using deep optical spectroscopic data (obtained with the 10-m
Keck telescope)$^{9}$ and ultra-deep X-ray observations (the 2 Ms
Chandra Deep Field-North; CDF-N)$^{18}$. As X-ray emission appears to
be an ubiquitous property of AGNs that is comparatively impervious to
obscuration, the combination of these deep datasets provides a
detailed census of AGN activity in SMGs. Here we focus on the role of
black-hole accretion in SMGs and the growth of black holes in massive
galaxies. The assumed cosmology is $H_0=65$~km~s$^{-1}$ Mpc$^{-1}$,
$\Omega_{\rm M}=1/3$, and $\Omega_{\Lambda}=2/3$.

The parent SMG sample comprises 20 submillimetre detected (with the
SCUBA camera on the James Clerk Maxwell Telescope) sources in the
CDF-N region whose positions have been reliably located from their
radio emission and which could then be spectroscopically
targeted$^{9}$. These sources have overall properties consistent with
the general SMG population ($S_{\rm 850\mu m}\approx$~4--12~mJy and
$z=$~0.56--2.91, with the majority at $z>2$)$^{6-9}$. Seventeen
($\approx$~85\%) of the 20 SMGs are detected at X-ray energies. A
detailed investigation of the nature of the X-ray emission has
revealed AGN activity in at least 15 ($\approx$~75\%) SMGs$^{19}$; the
properties of these sources are given in Table~1. The X-ray properties
of the other 5 ($\approx$~25\%) SMGs are consistent with those
expected from luminous star-formation activity, but we note that at
least one shows evidence for AGN activity at near-infrared
wavelengths$^{16}$.

\begin{figure}[t]
\centerline{\psfig{file=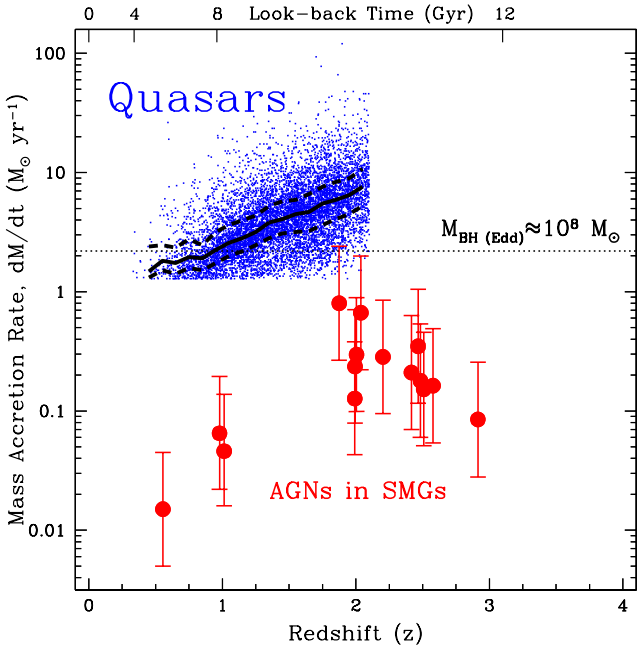,width=3.5in}}
\caption{Black-hole mass accretion rates. SMGs (filled circles), and a
  comparison sample of $M_{\rm B}<-24$ quasars (dots) are shown$^{4}$.
  The error bars represent the estimated uncertainty in the X-ray to
  bolometric correction for AGN activity in SMGs; see Table~1. The
  solid and dashed curves indicate the median and interquartile ranges
  for the comparison quasar sample. The dotted line indicates the
  approximate Eddington-limited mass accretion rate for an
  $\approx10^{8}$~$M_{\odot}$ black hole ($M_{\rm BH (Edd)}$). The
  SMGs have mass accretion rates approximately an order of magnitude
  lower than those of coeval quasars$^{4}$, suggesting that their
  black holes are smaller (typically $\approx
  10^{7}$--$10^{8}$~$M_{\odot}$); see text for additional black-hole
  mass constraints.}
\end{figure}

Two different observing modes were used in constructing the SMG
sample: 14 of the radio sources were specifically targeted with SCUBA
observations and 6 of the radio sources were detected in blank-field
SCUBA maps. Although targeting known radio sources with SCUBA
observations could potentially bias our AGN fraction owing to
contributions to the radio emission from AGN activity, a statistical
analysis does not reveal a strong AGN bias in our sample (P=1.0,
two-sided Fisher's exact test); see Table~1 for the observing modes of
the X-ray classified AGNs. However, due to the large positional
uncertainties of SCUBA sources ($\approx$~6--7 arcseconds), redshifts
could not be obtained for the $\approx$~35--50\% of the
radio-undetected SMG population, some of which might host AGN
activity$^{6,9}$. Making the conservative assumption that none of the
radio-undetected SMGs hosts AGN activity, our X-ray data alone suggest
an AGN fraction in the SMG population of $>38^{+12}_{-10}$\%. We note
that a similar fraction of comparably luminous galaxies in the local
Universe host AGN activity$^{22}$, although their space density is
approximately 3 orders of magnitude smaller than SMGs.

The large AGN fraction in the SMG population indicates that their
black holes are growing almost continuously throughout the intense
star-formation phase of these galaxies [i.e., assuming that all SMGs
are growing their black holes, the AGN duty cycle is
$>38^{+12}_{-10}$\%]. This suggests that the black holes and galaxy
spheroids are growing concordantly in SMGs$^{1}$. Such a close association
between AGN and star-formation activity is not seen in the coeval
field galaxy population or other coeval star-forming galaxy
populations ($\approx$~3--15\% AGN fraction)$^{23,24}$. The almost
continuous black-hole growth in SMGs suggests that there is an
abundance of available fuel, hinting that the accretion may be
occurring efficiently (i.e., at or close to the Eddington limit), as
predicted by theoretical studies of the growth of massive black
holes$^{25-27}$. The majority ($\approx$~85\%) of the AGNs are
obscured (see Table~1), as also predicted for black holes undergoing
efficient growth$^{25}$.

The unobscured X-ray luminosities of the AGNs ($L_{\rm
  X}\approx10^{43}$--$10^{44}$~erg~s$^{-1}$) suggest that the mass
accretion rates are modest ($\simlt$~1~$M_{\odot}$~yr$^{-1}$); see
Table~1 and Figure~1. By comparison to coeval quasars, the mass
accretion rates in SMGs are approximately an order of magnitude lower,
also suggesting modest black-hole masses (i.e.,
$\approx10^{7}$--$10^{8}$~$M_{\odot}$; see Table 1 and Figure 1).
Although these black-hole mass estimates are uncertain, the narrowness
of the broad emission lines (BELs; typical full-width half maximum
velocities of $\approx$~2500~km~s$^{-1}$)$^{16}$, when they are
visible, also suggest black-hole masses of $<10^{8}$~$M_{\odot}$ (ref
4). As well-studied AGNs with narrow BELs are found to be accreting at
(or even beyond) the Eddington limit$^{4}$, this provides further
support for the idea that the mass accretion rate in SMGs is
approximately Eddington limited. Under the assumption of
Eddington-limited accretion, the black holes in the SMGs will grow by
up to an order of magnitude over their $\approx10^{8}$~yr
star-formation lifetime. These overall properties are consistent with
those expected for massive galaxies undergoing rapid black-hole
growth.

The total black-hole growth from SMGs can be calculated by integrating
the average accretion density over the SMG redshift range; see
Figure~2.  Over the redshift interval $z=$~1.8--3.0 (corresponding to
80\% of the SMGs studied here), the black-hole density produced by
SMGs is $\approx6^{+11}_{-4}\times10^{3}$~$M_{\odot}$~Mpc$^{-3}$. To
put this quantity into context we need to compare it to the black-hole
growth from coeval quasar activity. Considering only quasars in the
luminosity range of $M_{\rm B}=$~--24 to --27 (the majority of which
reside in $\simgt$$M_{*}$ spheroid-dominated galaxies$^{28}$ and have
the properties consistent with being the progeny of SMGs$^{5}$), the
total SMG black-hole density is $\approx13^{+27}_{-9}$\% of the
black-hole growth from quasar activity$^{2}$. These constraints should
be considered lower limits since further pre-quasar growth at
$z=$~1.8--3.0 may be produced by radio-undetected SMGs and galaxies
with fainter submillimetre fluxes, if they host AGN activity. However, these
results are consistent with, for example, SMGs undergoing an intense
black-hole growth phase, where the black hole is grown from
$\approx10^{7}$ to $\approx10^{8}$~$M_{\odot}$, prior to a high
accretion-rate quasar phase, where the black hole is grown from
$\approx10^{8}$ to $\approx8\times10^{8}$ $M_{\odot}$. In this
scenario, although SMGs do not produce a large fraction of the
cosmological black-hole density, they are responsible for the crucial
pre-quasar phase where the black holes in massive galaxies are rapidly
grown. Indeed, the total black-hole growth from SMGs could only exceed
that from quasar activity if the quasar lifetime is insufficient to
double the mass of the black hole (i.e., $<3\times10^{7}$~yrs assuming
Eddington-limited accretion). Overall, this picture is in good
agreement with direct theoretical predictions of the black-hole growth
of SMGs and quasars$^{26,27}$.

What was the catalyst for the rapid black hole and stellar growth seen
in SMGs? Rest-frame ultra-violet images taken by the {\it Hubble Space
  Telescope} have shown that a considerably larger fraction of SMGs
are in major mergers (i.e., the merging of two galaxies of comparable
masses) than has been found in the coeval galaxy population$^{17,29}$.
Hydrodynamical simulations have shown that major mergers can
efficiently transport material toward the central regions of galaxies,
providing an effective mechanism to trigger nuclear star formation and
fuel the black hole$^{30}$. The result of these major mergers is
thought to be massive spheroid-dominated galaxies. Ultra-deep X-ray
observations of SMGs undergoing major mergers have shown that AGN
activity can be ongoing in both galactic components$^{12}$. Presumably
in these major-merger events the black holes will eventually coalesce,
further increasing the mass of the black hole in the resultant galaxy.

\begin{figure}[t]
\centerline{\psfig{file=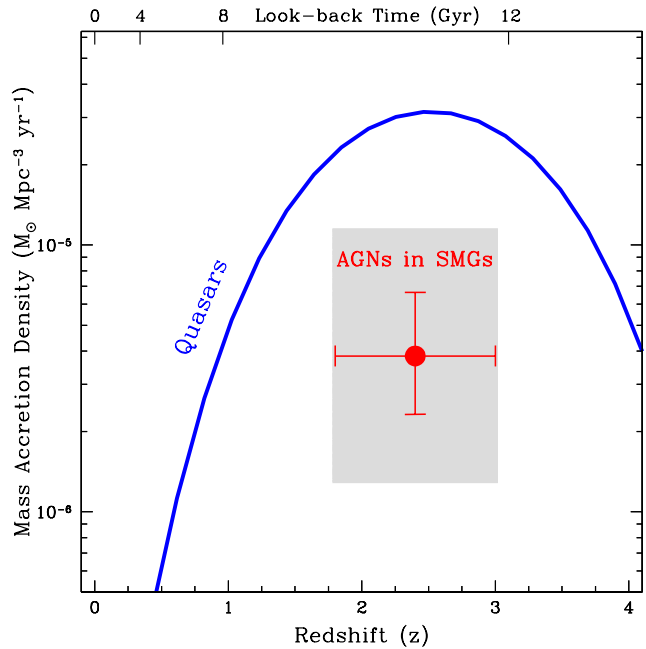,width=3.5in}}
\caption{Cosmological black-hole accretion density. SMGs (filled
circle), and $M_{\rm B}=$~--24 to --27 quasars (solid line)$^{2}$ are
shown. Error bars in the $x$-axis direction indicate the redshift
range and error bars in the $y$-axis direction represent $1\sigma$
Poisson counting uncertainties. The shaded region indicates the
uncertainty in the mass accretion rate conversion for the SMGs; see
Table~1. The mass accretion density from SMGs is
$\approx13^{+27}_{-9}$\% of that produced by coeval $M_{\rm B}=$~--24
to --27 quasar activity over the redshift interval $z=$~1.8--3.0.}
\end{figure}

%
%%%%%%%%%%%%%%%%%%%%%%%%%%%%%%%%%%%%%%%%%%%%%%%%%%%%%%%%%%%%%%%%%%%%%%
% TABLE
%%%%%%%%%%%%%%%%%%%%%%%%%%%%%%%%%%%%%%%%%%%%%%%%%%%%%%%%%%%%%%%%%%%%%%
%
\begin{table*}[!t]
\begin{center}
\normalsize
\caption{The Properties of the X-ray Classified AGNs}
\begin{tabular}{ccccccccc}
\hline
\\
  & $S_{\rm 850\mu m}$ &     &log$(L_{\rm X})$      & log($L_{\rm BOL})$  & ${dM}\over{dt}$        & $M_{\rm BH (Edd)}$ & X-ray & Observation\\
SMG name & (mJy)              & $z$ &(erg~s$^{-1}$)&($L_{\odot}$) & ($M_{\odot}$~yr$^{-1}$) & ($10^7$~$M_{\odot}$)    & Obscuration? & Mode? \\
\\
\hline
\\
123549.4 +621536 & 8.3$\pm$2.5 & 2.20 & 44.0 & 12.9 & $0.28^{+0.57}_{-0.19}$ & $1.3^{+2.6}_{-0.9}$ & Y & T \\
123555.1 +620901 & 5.4$\pm$1.9 & 1.88 & 44.4 & 13.2 & $0.80^{+1.60}_{-0.53}$ & $3.6^{+7.3}_{-2.4}$ & Y & T \\
123606.7 +621550 & 4.4$\pm$1.4 & 2.42 & 43.8 & 12.6 & $0.21^{+0.42}_{-0.14}$ & $1.0^{+1.9}_{-0.6}$ & N & M \\
123606.8 +621021 & 11.6$\pm$3.5 &2.51 & 43.7 & 13.1 & $0.15^{+0.31}_{-0.10}$ & $0.7^{+1.4}_{-0.5}$ & Y & T \\
123616.1 +621513 & 5.8$\pm$1.1 & 2.58 & 43.7 & 13.0 & $0.16^{+0.33}_{-0.11}$ & $0.7^{+1.5}_{-0.5}$ & Y & T \\
123622.6 +621629 & 7.7$\pm$1.3 & 2.47 & 44.0 & 13.1 & $0.35^{+0.70}_{-0.23}$ & $1.6^{+3.2}_{-1.1}$ & Y & T \\
123629.1 +621045 & 5.0$\pm$1.3 & 1.01 & 43.2 & 12.1 & $0.05^{+0.09}_{-0.03}$ & $0.2^{+0.4}_{-0.1}$ & Y & T \\
123632.6 +620800 & 5.5$\pm$1.3 & 1.99 & 43.9 & 12.9 & $0.24^{+0.47}_{-0.16}$ & $1.1^{+2.2}_{-0.7}$ & Y & M \\
123635.5 +621424 & 5.5$\pm$1.4 & 2.01 & 44.0 & 12.9 & $0.30^{+0.60}_{-0.20}$ & $1.4^{+2.7}_{-0.9}$ & Y & T \\
123636.7 +621156 & 7.0$\pm$2.1 & 0.56 & 42.7 & 11.1 & $0.02^{+0.03}_{-0.01}$ & $0.1^{+0.1}_{-0.1}$ & N & M \\
123707.2 +621408 & 4.7$\pm$1.5 & 2.48 & 43.8 & 12.9 & $0.18^{+0.36}_{-0.12}$ & $0.8^{+1.6}_{-0.5}$ & Y & T \\
123711.9 +621325 & 4.2$\pm$1.4 & 1.99 & 43.6 & 12.7 & $0.13^{+0.25}_{-0.08}$ & $0.6^{+1.2}_{-0.4}$ & Y & T \\
123712.0 +621212 & 8.0$\pm$1.8 & 2.91 & 43.4 & 12.7 & $0.09^{+0.17}_{-0.06}$ & $0.4^{+0.8}_{-0.3}$ & Y & M \\
123716.0 +620323 & 5.3$\pm$1.7 & 2.04 & 44.3 & 13.0 & $0.67^{+1.33}_{-0.44}$ & $3.0^{+6.1}_{-2.0}$ & Y & T \\
123721.8 +621035 & 12.0$\pm$3.9 &0.98 & 43.7 & 11.7 & $0.16^{+0.31}_{-0.10}$ & $0.7^{+1.4}_{-0.5}$ & Y & M \\
\\
\hline
\end{tabular}
\end{center}
\vskip 2pt
\footnotesize
{\sc Notes. ---} 
SMG name, submillimetre flux density ($S_{\rm 850\mu m}$), redshift (z), and
total bolometric luminosity ($L_{\rm BOL}$) are taken from ref 9. Unobscured
rest-frame 0.5--8.0 keV luminosity ($L_{\rm X}$), and the presence of
X-ray obscuration ($N_{\rm H}>10^{22}$~cm$^{-2}$) are taken from ref
19. The observation mode refers to whether the source was specifically
targeted with a SCUBA observation (``T'') or whether the source was
detected in a blank-field SCUBA map (``M''): 10 of the AGNs were
identified in targeted SCUBA observations and 5 of the AGNs were
identified in blank-field SCUBA maps. The mass accretion rate (dM/dt)
is estimated by converting the X-ray luminosity to the AGN bolometric
luminosity, under the assumption that the X-ray luminosity accounts
for $6^{+12}_{-4}$\% of the bolometric luminosity of the AGN (i.e.,
the range in conversion factors found in ref 20). Although our assumed
bolometric conversion factors were originally derived from unobscured
AGNs, while the majority of the AGNs in our sample are obscured, the
important issue is the relationship between the intrinsic X-ray
emission and the ultraviolet-optical accretion disk emission, which is
unlikely to be obscuration dependent. We also note that our bolometric
conversion factors agree with those estimated for obscured AGNs with
similar luminosities (and therefore similar mass accretion rates) to
the sources studied here$^{3,21}$. When converting from the bolometric
luminosity to the accreted mass we assumed the canonical efficiency of
10\% ($\epsilon=0.1$). The black-hole mass ($M_{\rm BH (Edd)}$) is
calculated from the mass accretion rate under the assumption that the
accretion is Eddington limited; see text for justification. The
black-hole masses will be higher if the accretion is sub Eddington;
see text for additional black-hole mass constraints.
\\
\end{table*}
\normalsize

%\bibliographystyle{nature}
%\bibliography{J1808burst}

\smallskip
\noindent {\small {\bf Acknowledgements.} We are grateful to Ross
McLure, Matt Page, Francesco Shankar, and Qingjuan Yu for providing
important data and useful scientific insight. We thank the Royal
Society (DMA; IS), PPARC (FEB), and NASA (SCC; WNB) for support. Data
presented herein were obtained using the W.M. Keck Observatory, which
is operated as a scientific partnership among Caltech, the University
of California and NASA.}

\medskip
\noindent {\small {\bf Competing interests statement.} The authors
  declare that they have no competing financial interests.}

\medskip
\noindent {\small {\bf Correspondence} and requests for materials should be
  addressed to DMA. (e-mail: dma@ast.cam.ac.uk).} 

\end{document}